# Neutral meson spectra measurements with ALICE

Yuri Kharlov, for the ALICE collaboration

*Institute for High Energy Physics, Protvino, 142281 Russia*

## Abstract

The ALICE experiment is equipped with electromagnetic calorimeters of two different types: PHOS, the lead-tungstate photon spectrometer, and EMCAL, the sampling lead-scintillator calorimeter. These two detectors measure photon spectra in a wide kinematic range and allow the reconstruction of neutral mesons decaying into photons. The measurement of the inclusive spectra of neutral meson production in pp collisions will not only provide a test of perturbative QCD but also important reference data for future heavy-ion measurements. The status of the analysis of the data accumulated during the LHC proton-proton runs at collision energies of 7 TeV and 900 GeV is shown, and the feasibility to measure the $\pi^0$ and $\eta$ meson spectra in proton-proton collisions is demonstrated.

## Introduction

Measurement of the inclusive hadron production at high transverse momentum remains a powerful tool to study perturbative QCD. The cross sections of these processes can be easily calculated, provided the parton density and fragmentation functions are known. Identification of charged hadrons is performed by measurements of the ionization loss, time of flight or velocity in the Cherenkov detectors, and is usually limited to a moderate $p_T$ range up to a few GeV/c. Instead, neutral meson spectra can be detected in the calorimeters via the photonic decays using invariant mass analysis. The $p_T$ range of these measurements is limited by the calorimeters granularity, and thus can be extended significantly in comparison with charged hadron measurements.

Neutral meson spectra measured in proton-proton collisions serve as a reference to the spectra measured in heavy ion collisions. Such measurements allowed the PHENIX experiment to perform extensive measurements on the nuclear modification factor $R_{AA}$ of $\pi^0$ and $\eta$ meson production in Au-Au collisions at $\sqrt{s_{NN}} = 200$ GeV in a wide $p_T$ range up to 20 GeV/c (1).

## ALICE setup

The ALICE experiment (2) is designed to study properties of hot quark matter at extreme energy density. The experiment comprises almost all possible types of detectors in order to fulfill comprehensive studies of nuclear matter via various signatures. Among the ALICE detectors, two electromagnetic calorimeters are installed to measure electromagnetic and hadronic probes by detection of photons and electrons.

One calorimeter, called EMCAL (ElectroMagnetic CALorimeter), is a sampling-type calorimeter consisting of lead-scintillator layers. The EMCAL will consist of 10 super-modules, 4 of which have been installed in 2009 before the first LHC run. Each super-module contains 1152 towers built into a matrix of 24x48 towers. The tower with a 6x6 cm$^2$ cross section contains 77 alternating layers of lead (1.4 mm thick) and plastic scintillator (1.7 mm thick). The acceptance of the EMCAL detector in the 2010 run was $|\eta| < 0.7$ in pseudorapidity and $80° < \varphi < 120°$ in the azimuthal angle. The material budget between the beam interaction point (IP) and the EMCAL front surface in terms of radiation length represents $0.8X_0$. The energy resolution is

$$\frac{\Delta E}{E} = \sqrt{\left(\frac{0.017 \text{ GeV}}{E}\right)^2 + \frac{(0.113 \text{ GeV}^{1/2})^2}{E} + 0.048^2}.$$

Another detector, photon spectrometer (PHOS), is a precise calorimeter built of inorganic scintillator crystals of lead tungstate (PbWO$_4$). The full PHOS geometry will contain 5 modules, out of which 3 modules have been installed in 2009 and took data in the first LHC run with proton beams in 2010. The modules consist of a matrix of 64x56 crystals with a 2.2x2.2 cm$^2$ cross section. The actual acceptance of the PHOS detector is $|\eta| < 0.13$ in pseudorapidity and $260° < \varphi < 320°$ in the azimuthal angle. The main advantages of the PHOS are a small material budget between the IP and the spectrometer surface, $0.2X_0$, and a high energy resolution

$$\frac{\Delta E}{E} = \sqrt{\left(\frac{0.013 \text{ GeV}}{E}\right)^2 + \frac{(0.033 \text{ GeV}^{1/2})^2}{E} + 0.011^2}.$$

## Data sample and event selection

The data of proton-proton collisions taken by the ALICE experiment in 2010 were taken with the minimum bias trigger. This trigger was based on the detectors surrounding the beam interaction point, the silicon pixel detector (SPD) and the scintillator detector V0 consisting of two rings located at the opposite sides from the interaction point, V0A and V0C. Any event selected by the beam crossing in the interaction point, which gave a signal in either SPD or V0A or V0C was triggered. The dead time of the data acquisition after triggering the proton-proton collision, was determined by the slowest ALICE detector and was about 1 µs.

The current analysis of neutral meson measurements is based on the data taken in May-August 2010. Proton-proton collisions at $\sqrt{s} = 900$ GeV were delivered by the LHC in May 2010 with the total statistics of $6.1 \cdot 10^6$ events which corresponds to the integrated luminosity $\int \mathcal{L}dt = 0.12$ nb$^{-1}$. The total statistics of $pp$ collisions at $\sqrt{s} = 7$ TeV taken in June-August 2010 is $2.1 \cdot 10^8$ events with the integrated luminosity $\int \mathcal{L}dt = 3$ nb$^{-1}$.

Expected differential cross sections of the $\pi^0$ and $\eta$ meson production have been estimated in perturbatve QCD (3), using next-to-leading order calculations with the INCNLO code (4) for the $\pi^0$ and leading order calculations with Pythia (5) for the $\eta$ meson. According to these predictions, the $\pi^0$ production spectrum in $pp$ collisions at $\sqrt{s} = 7$ TeV at the integrated luminosity $\int \mathcal{L}dt = 10$ nb$^{-1}$ can be measured in ALICE up to $p_T$<25 GeV/c, and the $\eta$ meson production spectrum is feasible at $p_T$<20 GeV/c.

## Reconstruction and analysis

Raw data taken by the ALICE data acquisition system during the LHC runs, were recorded to the GRID mass storage of the Tier-0 center, and was reconstructed immediately after the end of the run, as soon as data migration has been completed to GRID. The reconstructed data was stored in the ESD format (event summary data) which contains all essential information needed for the physics analysis. The ALICE calorimeters, EMCAL and PHOS, stored to the ESD clusters, as well as the amplitudes of the individual cells. The ESD clusters can be considered as candidates for physics particles detected by the calorimeters. They are characterized by the energy and the coordinate in the global reference system, the 4-momentum of the photon candidate associated with the cluster, and various shower shape parameters needed for the particle identification.

In the reconstruction procedure, the measured amplitudes of the calorimeter cells are transformed to the deposited energy using the calibration parameters. The EMCAL detector was pre-calibrated by cosmic rays before installation to the ALICE setup. PHOS was installed without such pre-calibration. Upon installation, the high-voltage bias of the photo-detectors – avalanche photodiodes – was set for both calorimeters to provide the same gains, according to the APD datasheets provided by the vendor, Hamamatsu. This adjustment of the APD gains was sufficient only for a 20-50% relative calibration. Further improvement of calibration parameters was performed using the physics data of proton-proton collisions. Equalization of the mean deposited energy in each cell was used for relative calibration of the PHOS detector, then adjusting the measured $\pi^0$ mass to the PDG value gave the absolute calibration. The relative and absolute calibration of EMCAL was found via equalization of the $\pi^0$ peak position measured in each EMCAL cell. The accuracy of the found calibration parameters for both calorimeters was estimated to be at the level of 6.5-7%.

To minimize a possible bias of the particle identification, any clusters in the EMCAL and PHOS with the energy above the minimum ionizing signal, $E > 0.3$ GeV, and containing at least 2 cells in EMCAL and 3 cells in PHOS, were considered as photon candidates. Invariant mass spectra of the photon candidate pairs $M_{\gamma\gamma}$ were evaluated in different bins of transverse momentum $p_T$ of the pair. Examples of the invariant mass spectra in some selected $p_T$ intervals in EMCAL and PHOS are shown in Figs.1 in the mass range of the $\pi^0$ and in Fig.2 in the mass range of the $\eta$ meson. The $\pi^0$ and $\eta$ meson raw spectra, or the number of reconstructed mesons as a function of $p_T$, were obtained from the photon pair invariant mass distributions by extraction the signal in the resonance peaks above the background at the $\pi^0$ and $\eta$ masses. The signal was extracted via fitting the mass spectra by the sum of a Gaussian describing the resonance peak and a polynomial for the combinatorial background. The fitting function is represented by the curves in Figs.1-2. The raw spectra of $\pi^0$ and $\eta$ mesons detected in $pp$ collisions in EMCAL and PHOS are shown in Figs.3-6. The analyzed statistics allows to measure the $\pi^0$ spectrum at $0.5 < p_T < 20$ GeV/c in $pp$ collisions at $\sqrt{s} = 7$ TeV and at $0.5 < p_T < 5$ GeV/c at $\sqrt{s} = 900$ GeV. The $\eta$ meson spectrum can be measured at $\sqrt{s} = 7$ TeV in the range $5 < p_T < 11$ GeV/c in PHOS and $2 < p_T < 18$ GeV/c in EMCAL.

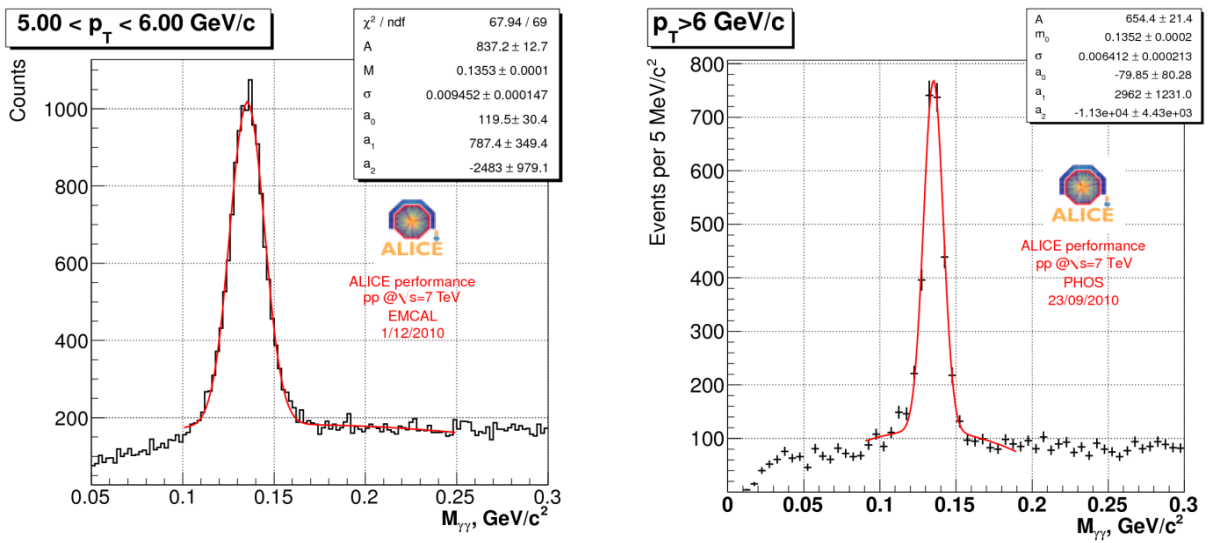

Fig.1. Invariant mass spectra around $\pi^0$ in EMCAL (left) and in PHOS (right) in $pp$ collisions at $\sqrt{s} = 7$ TeV.

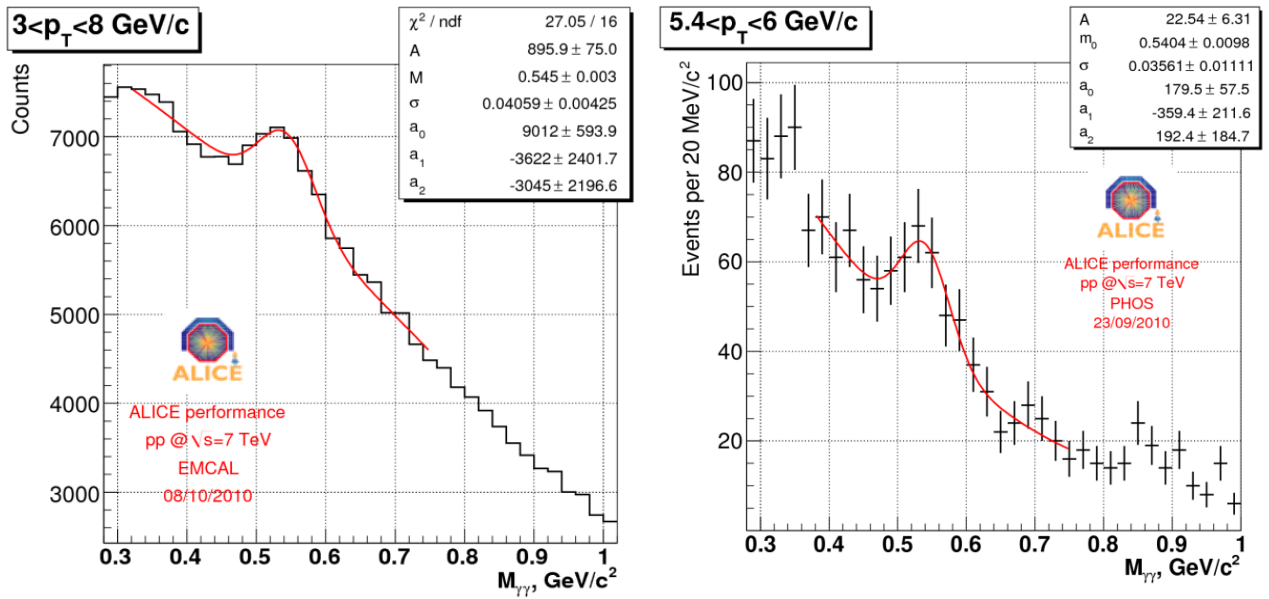

Fig.2. Invariant mass spectra around $\eta$ in EMCAL (left) and in PHOS (right) in $pp$ collisions at $\sqrt{s} = 7$ TeV.

The systematic uncertainties of the measured spectra have various sources. The finite accuracy of the calorimeter calibration leads to a systematic uncertainty in the measured spectra of 5%. Unknown non-linearity of the energy scale of the calorimeters is significant at low transverse momentum, $p_T < 1.5$ GeV/c, where it was estimated to be 15%, while at higher $p_T$ its contribution is negligible, less than 2%. Uncertainty in the geometrical acceptance was found to be less then 2%. A similar contribution comes from uncertainties in the photon conversion loss in the ALICE medium, in the raw spectrum extraction and in the particle identification. The total systematic uncertainty at $p_T > 1.5$ GeV/c was estimated to be about 10%.

## Conclusion

The ALICE electromagnetic calorimeters, EMCAL and PHOS, have collected significant statistics in proton-proton collisions at the LHC in 2010 which allow to measure the production spectra of $\pi^0$ and $\eta$ mesons at mid-rapidity in a wide $p_T$ range. The $\pi^0$ spectrum can be measured at $0.5 < p_T < 20$ GeV/c in $pp$ collisions at $\sqrt{s} = 7$ TeV and at $0.5 < p_T < 5$ GeV/c at $\sqrt{s} = 900$ GeV. The $\eta$ meson spectrum can be measured in $pp$ collisions at $\sqrt{s} = 7$ TeV in the range $2 < p_T < 18$ GeV/c. The presented analysis has demonstrated the feasibility of such measurements. These measurements allow to test the perturbative QCD predictions and provide reference data for the heavy ion collisions.

This work was supported partially by the RFBR grant 10-02-91052.

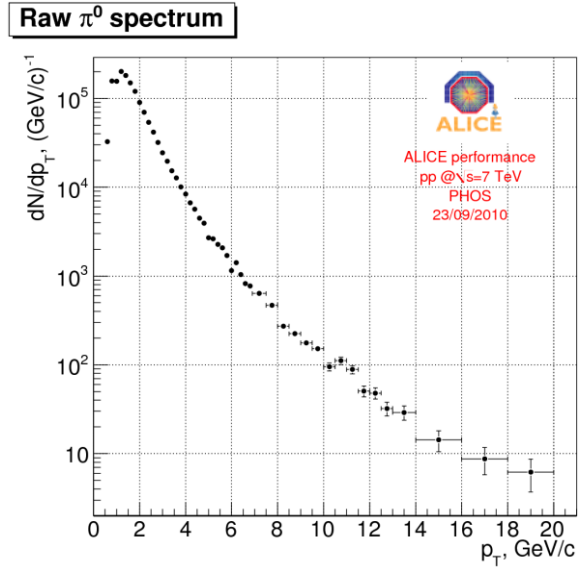

Fig.3. Reconstructed $\pi^0$ raw spectrum in PHOS in $pp$ collisions at $\sqrt{s} = 7$ TeV.

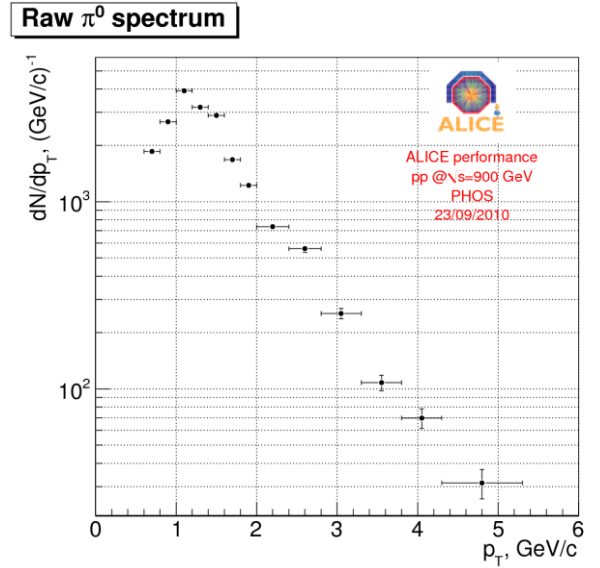

Fig.4. Reconstructed $\pi^0$ raw spectrum in PHOS in $pp$ collisions at $\sqrt{s} = 900$ GeV.

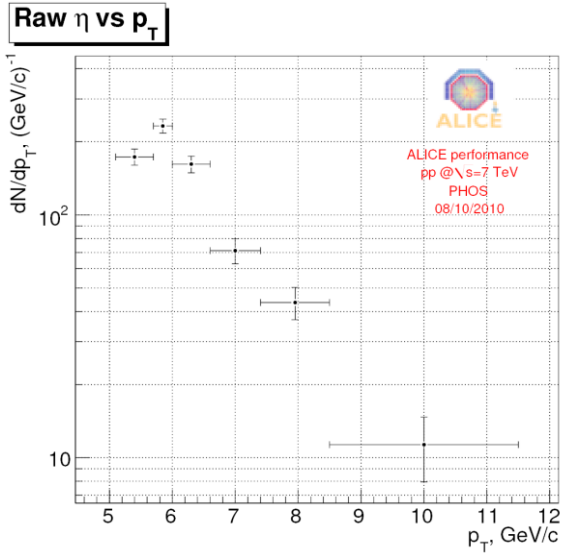

Fig.5. Reconstructed $\eta$ raw spectrum in PHOS in $pp$ collisions at $\sqrt{s} = 7$ TeV.

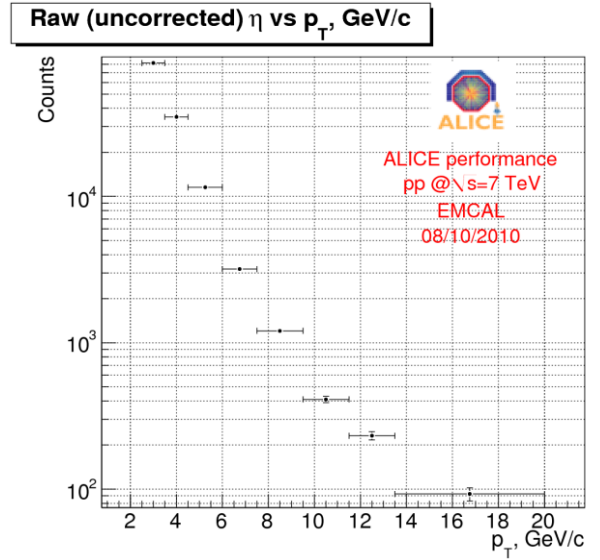

Fig.6. Reconstructed $\eta$ raw spectrum in EMCAL in $pp$ collisions at $\sqrt{s} = 7$ TeV.